\documentclass[12pt,preprint,showpacs,preprintnumbers]{revtex4}
\usepackage{amsfonts}
\usepackage{amsmath}
\usepackage{amssymb}
\usepackage{hyperref}
\usepackage{color}
\usepackage{graphicx}
\usepackage{amssymb}
\usepackage{amsmath}
\usepackage{graphicx}
\usepackage{dcolumn}
\usepackage{bm}
\usepackage{epsfig}
\usepackage[T1]{fontenc}
\usepackage{ae,aecompl}

\begin{document}

\title{Discrete Boltzmann modeling of liquid-vapor system}
\author{ Aiguo Xu$^{1,2}$\thanks{%
Corresponding author: Xu\_Aiguo@iapcm.ac.cn}, Guangcai Zhang$^{1}$, Yanbiao
Gan$^{3}$}
\affiliation{$^1$ National Key Laboratory of Computational Physics, Institute of Applied
Physics and Computational Mathematics, P. O. Box 8009-26, Beijing 100088,
P.R.China \\
$^2$ Center for Applied Physics and Technology, MOE Key Center for High
Energy Density Physics Simulations, College of Engineering, Peking
University, Beijing 100871, China \\
$^3$ North China Institute of Aerospace Engineering, Langfang 065000,
P.R.China }
\date{\today }

\begin{abstract}
We further probe the Discrete Boltzmann Modeling(DBM) of the
single-component two phase flows or the liquid-vapor system. There are two
kinds of nonequilibrium effects in the system. The first is the Mechanical
NonEquilibrium(MNE). The second is the Thermodynamic NonEquilibrium(TNE).
The MNE is well described in the traditional fluid dynamic theory. The
description of TNE resorts to the gas kinetic theory. Since based on the
Boltzmann equation, the DBM makes possible to analyze both the MNE and TNE.
The TNE is the main discussion of this work. A major purpose of this work is
to show that the DBM results can be used to confirm and/or improve the
macroscopic modeling of complex system.
\end{abstract}

\pacs{47.11.-j, 47.40.-x, 47.55.-t, 05.20.Dd}
\maketitle

\section{Introduction}

\subsection{Calssification of LB methods}

In the past two decades the Lattice Boltzmann (LB) method has been becoming
a powerful simulation tool in various complex fluids\cite{Succi-Book},
especially in multiphase flows, a few examples are referred to Refs.\cite%
{Shan-Chen,Swift,XGL2003,Xu2005,XGL2005,XGL2006,XuGan2011PRE,XuGan2012EPL,XuGan2012FoP,Yeomans}%
. In most of the existing studies on LB methods and LB simulations, the LB
generally appears as a kind of new numerical scheme to solve the
corresponding partial differential equations. As pointed out in the early
studies\cite{Succi-Book}, appropriately designed LB can work as a new kind
of mesoscopic kinetic model for various complex behaviors\cite%
{XuReview2012,XuReview2014}. In this work we refer a LB model in the second
class to as a Discrete Boltzmann Model(DBM). Besides recovering the
macroscopic hydrodynamic equations in the continuum limit, a DBM should
present more kinetic information on the nonequilibrium effects which are
generally related to some mesoscopic structures.

\subsection{Classification of nonequilibrium behaviors}

For a flow system without chemical reaction, there are two kinds of
nonequilibrium behaviors, the Mechanical NonEquilibrium (MNE) and
Thermodynamic NonEquilibrium (TNE). The MNE makes the acceleration, breaks
the steady state, and consequently results in the further evolution of the
flow system. The MNE is generally well described by the traditional fluid
mechanics which is based on the continuous assumption and describes the
evolution of macroscopic quantities like the density $\rho (\mathbf{r},t)$,
the momentum $\rho (\mathbf{r},t)\mathbf{u}(\mathbf{r},t)$ and the energy $E(%
\mathbf{r},t)$, where $\mathbf{r}$ and $t$ are the position and time. The
evolution equations of the traditional fluid mechanics appear also as the
conservation laws of the mass, momentum and energy. The theory at this level
do not access the behavior of the molecules. As the fundamental equation of
nonequilibrium statistical physics, the Boltzmann equation describes the
flow system from a more fundamental level. It resorts to the concept of
distribution function $f(\mathbf{r},\mathbf{v},t)$, where $\mathbf{v}$ is
the molecular velocity. In the thermodynamic equilibrium state, the
molecular velocity follows the Maxwellian distribution function $f^{(0)}(%
\mathbf{r},\mathbf{v},t)$. The Boltzmann equation with the BGK model reads,

\begin{equation}
\frac{\partial f}{\partial t}+v_{\alpha }\frac{\partial f}{\partial
r_{\alpha }}+a_{\alpha }\frac{\partial f}{\partial v_{\alpha }}=-\frac{1}{%
\tau }\left( f-f^{(0)}\right) \text{,}  \label{eq1}
\end{equation}%
where $\tau $ is the relaxation time, $\mathbf{a}$ is an external force
acting on the molecule with the unit mass, and $\alpha =x$, $y$. The first
three low order moments of the distribution function $f$ are related to the
density $\rho $, momentum $\rho \mathbf{u}$ and energy $E\mathbf{=}\frac{1}{2%
}\rho \mathbf{u}^{2}+\rho T$ as below:

\begin{equation*}
\int fd\mathbf{v=}\int f^{(0)}d\mathbf{v=}\rho ,
\end{equation*}%
\begin{equation*}
\int \mathbf{v}fd\mathbf{v=}\int \mathbf{v}f^{(0)}d\mathbf{v=}\rho \mathbf{%
u,}
\end{equation*}%
\begin{equation*}
\int \frac{1}{2}\mathbf{v}^{2}fd\mathbf{v=}\int \frac{1}{2}\mathbf{v}%
^{2}f^{(0)}d\mathbf{v}=\frac{1}{2}\rho \mathbf{u}^{2}+\rho T,
\end{equation*}%
where $\mathbf{u}$ and $T$ are the local flow velocity and temperature,
respectively. For the convenience of description, we use $\mathbf{M}%
_{m,n}(f) $ to denote the moment of $f$. It is a $n$-th order tensor in
molecule velocity $\mathbf{v}$ contracted from the $m$-th tensor. It is
clear that $\mathbf{M}_{0,0}(f)=\mathbf{M}_{0,0}(f^{(0)})=\rho $, $\mathbf{M}%
_{1,1}(f)=\mathbf{M}_{1,1}(f^{(0)})=\rho \mathbf{u}$, $\mathbf{M}_{2,2}(f)=%
\mathbf{M}_{2,2}(f^{(0)})=E$\textbf{. }The higher order moments are related
to macroscopic quantities which are not neccessarily conserved during the
evolution. Besides $\rho $, $\rho \mathbf{u}$ and $E$, the Boltzmann
equation is also capble of accessing non-conserved macroscopic quantities,
which is the point being beyond the traditional fluid mechanics. The
difference of a non-conserved macroscopic quantity, $\mathbf{M}_{m,n}(f)$,
from its value in corresponding thermodynamic equilibrium,$\mathbf{M}%
_{m,n}(f^{(0)})$, is a kind of manifestation or measure of the TNE. For the
convenience of describing the pure TNE effects, we define also the central
moment $\mathbf{M}_{m,n}^{\ast }(f)$ which is the $n$-th order tensor in
relative velocity $(\mathbf{v}-\mathbf{u)}$ contracted from the $m$-th
tensor.

\subsection{Progress of discrete Boltzmann modeling TNE}

An appropriately designed DBM should inherit partly this merit\cite%
{XuReview2012,XuReview2014}.

The idea using DBM to access the TNE behavior\cite{XuReview2012} has been
further specified and applied in various compressible flow systems via
several models\cite%
{XuYan2013,XuGan2013,XuLin2014PRE,XuLin2014arXiv,XuChen2014FoP}. Examples of
LBGK for compressible flow systems are referred to Refs. \cite%
{XuGan2013,XuLin2014PRE} where preliminary studies on shocking behavior are
shown. An example of MRT-LB for compressible flow systems is referred to
Ref. \cite{XuChen2014FoP}. In Refs.\cite{XuYan2013,XuLin2014arXiv} the TNE
behaviors in combustion systems are initially investigated via LBGK models.
Up to now, all the previous studies on the discrete Boltzmann modeling of
TNE behaviors were for systems following the ideal-gas equation of state.

In this work we further probe the kinetic nature of DBM for the non-ideal
gas systems, particularly, the liquid-vapor system or single-component two
phase flows.

\section{Kinetic modeling of non-ideal gas system}

For the ideal gas system, $\mathbf{a}=0$, and the local pressure $P=\rho T$.
If do not consider the boundary effects, the gradient of any macroscopic
quantity, for example $\nabla \rho $, $\nabla \mathbf{u}$, $\nabla T$, or $%
\nabla P$, work as a driving force for thermodynamic nonequilibrium effects.
For the liquid-vapor system, the existence of the interparticle force makes
the situation more complex. The force term in the LB equation is a second
kind of driving force for the TNE effects. The two kinds of driving forces
tend to balance each other. The competition of the two kinds of driving
forces determines the evolution of the system. When the system arrives at
its steady state, i.e., the mechanical equilibrium, the two kinds of driving
forces balance each other.

\subsection{Maxwellian or not in steady state?}

A question here is whether or not the distribution function is the
Maxwellian when the liquid-vapor system is in a steady state. We first
assume and confirm it is.

In the steady state, if
\begin{equation}
f=f^{(0)}\text{,}  \label{eq5}
\end{equation}%
then, equation (\ref{eq1}) requires
\begin{equation}
v_{\alpha }\frac{\partial f^{(0)}}{\partial r_{\alpha }}+a_{\alpha }\frac{%
\partial f^{(0)}}{\partial v_{\alpha }}=0\text{.}  \label{eq6}
\end{equation}%
When the inhomogeneity exists, for example, around the interfaces,
\begin{eqnarray}
\frac{\partial f^{(0)}}{\partial r_{\alpha }} &\neq &0\text{,}  \label{eq7}
\\
\frac{\partial f^{(0)}}{\partial v_{\alpha }} &\neq &0\text{,}  \label{eq8}
\end{eqnarray}%
the effects of the force term and the advection term balance each other.
When the inhomogeneity does not exists, for example, far from the interfaces,

\begin{eqnarray}
\frac{\partial f^{(0)}}{\partial r_{\alpha }} &=&0\text{,}  \label{eq9} \\
\frac{\partial f^{(0)}}{\partial v_{\alpha }} &\neq &0\text{.}  \label{eq10}
\end{eqnarray}%
It is clear that the advection term makes no effects. Now, we check the
force term in particle velocity derivative. In the currently using
liquid-vapors models, the force $a_{\alpha }$ is from the spatial derivative
of some macroscopic quantity like the pressure or density. Therefore, $%
a_{\alpha }=0$ in the homogeneous region. Up to this step, we can confirm
physically that, in the steady state, $f=f^{(0)}$.

\subsection{Where are the meaningful nonequilibrium effects?}

In the steady state, no matter far from or around the interfaces, $f=f^{(0)}$%
. Consequently, the total or net TNE effects
\begin{eqnarray}
\boldsymbol{\Delta }_{m,n} &=&\mathbf{M}_{m,n}(f)-\mathbf{M}_{m,n}(f^{(0)})=%
\mathbf{0.}  \label{eq11a} \\
\boldsymbol{\Delta }_{m,n}^{\ast } &=&\mathbf{M}_{m,n}^{\ast }(f)-\mathbf{M}%
_{m,n}^{\ast }(f^{(0)})=\mathbf{0.}  \label{eq11b}
\end{eqnarray}%
But before arriving at the steady state, $\boldsymbol{\Delta }_{m,n}$ and $%
\boldsymbol{\Delta }_{m,n}^{\ast }$ can work as measures of the TNE.

The kinetic model (\ref{eq1}) can also be rewriiten as%
\begin{equation}
\frac{\partial f}{\partial t}+v_{\alpha }\frac{\partial f}{\partial
r_{\alpha }}=-\frac{1}{\tau }\left( f-f^{(0)NEW}\right)   \label{eq14}
\end{equation}%
where
\begin{equation}
f^{(0)NEW}=f^{(0)}-\tau a_{\alpha }\frac{\partial f}{\partial v_{\alpha }}
\label{eq15}
\end{equation}%
can be regarded as a new equilibrium state shifted due to the exsitence of
interparticle interactions. We define%
\begin{eqnarray}
\boldsymbol{\Delta }_{m,n}^{C} &=&\mathbf{M}_{m,n}(f)-\mathbf{M}%
_{m,n}(f^{(0)NEW}),  \label{eq15b} \\
\boldsymbol{\Delta }_{m,n}^{\ast C} &=&\mathbf{M}_{m,n}^{\ast }(f)-\mathbf{M}%
_{m,n}^{\ast }(f^{(0)NEW}).  \label{eq15c}
\end{eqnarray}%
In the steady state,
\begin{eqnarray}
\boldsymbol{\Delta }_{m,n}^{C} &=&\mathbf{M}_{m,n}(f-f^{(0)}+\tau a_{\alpha }%
\frac{\partial f}{\partial v_{\alpha }})  \label{eq16} \\
&=&\mathbf{M}_{m,n}(\tau a_{\alpha }\frac{\partial f}{\partial v_{\alpha }})
\label{eq17} \\
&\approx &\mathbf{M}_{m,n}(\tau a_{\alpha }\frac{\partial f^{(0)}}{\partial
v_{\alpha }})  \label{eq18} \\
&=&\mathbf{M}_{m,n}(-\tau \frac{a_{\alpha }}{T}(v_{\alpha }-u_{\alpha
})f^{(0)})\text{,}  \label{eq19}
\end{eqnarray}%
\begin{eqnarray*}
\boldsymbol{\Delta }_{m,n}^{\ast C} &=&\mathbf{M}_{m,n}^{\ast
}(f-f^{(0)}+\tau a_{\alpha }\frac{\partial f}{\partial v_{\alpha }}) \\
&\approx &\mathbf{M}_{m,n}^{\ast }(-\tau \frac{a_{\alpha }}{T}(v_{\alpha
}-u_{\alpha })f^{(0)})\text{.}
\end{eqnarray*}%
It is clear that in the steady state $\boldsymbol{\Delta }_{m,n}^{C}$ or $%
\boldsymbol{\Delta }_{m,n}^{\ast C}$ (approximately) describes the TNE due
to the interparticle interactions or interfacial forces. Before arriving at
the steady state, it (approximately) describes the combined effects of the
inhomogeneities and the interfacial forces.

\subsection{Beyond the traditional fluid dynamics}

In most studies based on the traditional fluid dynamics, the heat conduction
is assumed to follow the Fourier law,
\begin{equation}
q=\lambda \nabla T\text{,}  \label{eq12}
\end{equation}%
and the viscous stress is assumed to follow the Newton model,

\begin{equation}
\boldsymbol{\sigma }=\gamma \nabla \mathbf{u}\text{,}  \label{eq13}
\end{equation}%
where the coefficients $\lambda $ and $\gamma $ are constants. The DBM
should be capable of presenting results for checking the validity of the two
assumptions and to find more reasonable responsive relations for the complex
transportation processes.

The steady state is a trivial case for the LB simulation. The main purpose
of the DBM is to simualte the evolution of the system before it arrives at
the final steady state, and the DBM results should confirm or being helpful
for improving the physical modeling from the macroscopic scale.

The relations $\boldsymbol{\Delta }^{*}_{m,n}$ versus the gradients of
macroscopic quantities ($\nabla \rho $, $\nabla T$, $\nabla \mathbf{u}$, $%
\nabla P$) show the nonequilibrium responses to nonequilibrium driving
forces. From these results, the Fourier law, Newton assumption, etc., can be
checked, and improved and more interesting dependences may be found.

\section{Discrete Kinetic model}

In constructing the discrete Boltzmann kinetic models for non-ideal gas
system, a key step is to approximate $f$ by $f^{(0)}$ in the force term of
the original BGK equation before discretizing the space of particle
velocity. Since%
\begin{equation}
\frac{\partial f}{\partial v_{\alpha }}\approx \frac{\partial f^{(0)}}{%
\partial v_{\alpha }}=\frac{1}{T}(v_{\alpha }-u_{\alpha })f^{(0)}
\label{eq20}
\end{equation}%
Therefore, the currently using DBM for liquid-vapor system reads,%
\begin{equation}
\frac{\partial f_{j}}{\partial t}+v_{j\alpha }\frac{\partial f_{j}}{\partial
r_{\alpha }}+\frac{a_{\alpha }}{T}(v_{j\alpha }-u_{\alpha })f_{j}^{(0)}=-%
\frac{1}{\tau }\left( f_{j}-f_{j}^{(0)}\right) ,  \label{eq21}
\end{equation}%
or equivalently,
\begin{equation}
\frac{\partial f_{j}}{\partial t}+v_{j\alpha }\frac{\partial f_{j}}{\partial
r_{\alpha }}=-\frac{1}{\tau }\left[ f_{j}-f_{j}^{(0)NEW}\right]  \label{eq23}
\end{equation}%
where the subscript \textquotedblleft $j$\textquotedblright\ is the index of
the discrete velocity and
\begin{equation}
f_{j}^{(0)NEW}=f_{j}^{(0)}+\tau \frac{a_{\alpha }}{T}(v_{\alpha }-u_{\alpha
})f_{j}^{(0)}\text{.}  \label{eq24}
\end{equation}%
In the steady state,%
\begin{equation}
\boldsymbol{\Delta }_{m,n}^{C}=\mathbf{M}_{m,n}(-\tau \frac{a_{\alpha }}{T}%
(v_{\alpha }-u_{\alpha })f_{j}^{(0)})\text{.}  \label{eq25}
\end{equation}%
\begin{equation}
\boldsymbol{\Delta }_{m,n}^{\ast C}=\mathbf{M}_{m,n}^{\ast }(-\tau \frac{%
a_{\alpha }}{T}(v_{\alpha }-u_{\alpha })f_{j}^{(0)})\text{.}
\end{equation}%
Under the framework of DBM, in the definitions of $\mathbf{M}_{m,n}$ and $%
\mathbf{M}_{m,n}^{\ast }$, $\int d\mathbf{v}$ is replaced by $%
\sum\limits_{j}$.

\section{Conclusion}

We present a framework for constructing discrete Boltzmann Kinetic model for
the liquid-vapor system or single-component two phase flows. Besides the
mechanical nonequilibrium being well described by traditional computational
fluid dynamics, the discrete Boltzmann kinetic model can be applied to
access the thermodynamic nonequilibrium behaviors. The idea presented here
works also for constructing MRT-DBM for non-ideal gas systems.

\end{document}